\documentclass[twocolumn,showpacs,preprintnumbers,amsmath,amssymb]{revtex4}

\usepackage{graphicx}
\usepackage{dcolumn}
\usepackage{bm}

\begin{document}

\preprint{Submission to Phys. Rev. B}

\title{
Magnetic structure of 
the spin-1/2 frustrated 
quasi-one-dimensional antiferromagnet Cu$_3$Mo$_2$O$_9$:  
Appearance of a partial disordered state
}

\author{Masashi Hase$^1$}
 \email{HASE.Masashi@nims.go.jp}
\author{Haruhiko Kuroe$^2$}
\author{Vladimir Yu. Pomjakushin$^3$}
\author{Lukas Keller$^3$}
\author{Ryo Tamura$^1$}
\author{Noriki Terada$^1$}
\author{Yoshitaka Matsushita$^1$}
\author{Andreas D\"onni$^1$}
\author{Tomoyuki Sekine$^2$}

\affiliation{%
${}^{1}$National Institute for Materials Science (NIMS), 
1-2-1 Sengen, Tsukuba, Ibaraki 305-0047, Japan \\
${}^{2}$Department of Physics, Sophia University, 
7-1 Kioi-cho, Chiyoda-ku, Tokyo 102-8554, Japan \\
${}^{3}$Laboratory for Neutron Scattering and Imaging, Paul Scherrer Institut (PSI), 
CH-5232 Villigen PSI, Switzerland
}%

\date{\today}

\begin{abstract}

We investigated the crystal and magnetic structures of  
the spin-1/2 frustrated antiferromagnet Cu$_3$Mo$_2$O$_9$ 
in which the spin system consists of antiferromagnetic chains and dimers.  
The space group at room temperature 
has been reported to be orthorhombic $Pnma$ (No. 62). 
We infer that 
the space group above $T_{\rm N} = 7.9$ K is monoclinic $P2_1/m$ (No. 11)
from the observation of reflections forbidden in $Pnma$ 
in x-ray powder diffraction experiments at room temperature. 
We determined the magnetic structure of Cu$_3$Mo$_2$O$_9$
in neutron powder diffraction experiments.  
Magnetic moments on dimer sites lie in the $ac$ planes. 
The magnitudes are $0.50 \sim 0.74 \mu_{\rm B}$. 
Moments on chain sites may exist 
but the magnitudes are very small. 
The magnetic structure indicates that 
a partial disordered state is realized.  
We consider the origin of the magnetic structure, weak ferromagnetism, 
and electric polarization. 

\end{abstract}


\maketitle

\section{INTRODUCTION}

Several frustrated antiferromagnets exhibit intriguing magnetic states such as 
the quantum spin-liquid state \cite{Anderson73}, 
the chiral ordered state \cite{Miyashita84,Onoda07}, 
the spin nematic or the multipolar state \cite{Tsunetsugu06,Zhitomirsky08}, 
and the spin-gel state \cite{Kawamura10}. 
Among frustrated antiferromagnets, 
frustrated spin chains provide grounds for 
studies of exotic quantum phases caused by 
combination of frustration and quantum fluctuation. 
In the Heisenberg spin-1/2 chain with antiferromagnetic (AF)
nearest-neighbor (NN) and next-nearest-neighbor (NNN) exchange interactions
($J_{\rm NN}$ and $J_{\rm NNN}$ interactions, respectively), 
the ground state(s) are 
a gapless spin-singlet state designated as Tomonaga-Luttinger liquid and 
two-folded-degenerated gapped spin-singlet states 
for $0 \le J_{\rm NNN} < 0.241 J_{\rm NN}$ and 
$0.241 J_{\rm NN} < J_{\rm NNN}$, respectively 
\cite{Haldane82a,Haldane82b,Okamoto92}. 
In the spin-Peierls substance CuGeO$_3$ \cite{Hase93a,Hase93b,Hase93c}, 
both the interactions are considered to exist \cite{Lorenzo94,Castilla95,Riera95}. 
It is inferred that 
a large spin gap observed in Raman scattering experiments of 
3.5 \% Mg-doped CuGeO$_3$ under high pressures 
is generated 
not only by the spin-Peierls transition 
but also by the frustration between the two interactions \cite{Tanokura03}. 
In the spin-1/2 chain with 
ferromagnetic $J_{\rm NN}$ and AF $J_{\rm NNN}$ interactions 
in the presence of magnetic fields, 
theoretical studies predict various quantum phases including 
the vector chiral phase, the spin nematic phase, phases with multipole order, 
and the spin-density-wave phases 
\cite{Chubukov91,Kolezhuk05,Heidrich06,Vekua07,Kecke07,Hikihara08,Sudan09,Sato09,Heidrich09,Sato11}.  
Several model substances have been found and are summarized 
in Table 1 in Ref.~\cite{Hase04} or Fig. 5 in Ref.~\cite{Drechsler07}. 
Experimental results are compared with theoretical results \cite{Hikihara08,Sato11}.
In the frustrated diamond-chain system, 
three ground states appear according to 
relative intensities among exchange interactions \cite{Okamoto99}. 
A spin-liquid ground state and a 1/3 magnetization plateau were found in the model compound Cu$_3$(CO$_3$)$_2$(OH)$_2$ \cite{Kikuchi05}. 

Cu$_3$Mo$_2$O$_9$ provides a different frustrated spin-1/2 AF 
chain \cite{Hamasaki08}. 
Figure 1 shows schematically the spin system in Cu$_3$Mo$_2$O$_9$. 
Here, we assume that 
the space group SG is orthorhombic $Pnma$ (No. 62) 
\cite{Steiner97,Reichelt05}.
There are three crystallographic Cu$^{2+}$ sites having spin-1/2. 
Four types of NN exchange interactions between Cu spins exist. 
The $J_4$ interaction forms AF chains 
parallel to the $b$ axis of Cu1 spins. 
The $J_3$ interaction forms AF dimers in $ac$ planes 
of Cu2 and Cu3 spins. 
The $J_1$ and $J_2$ interactions connect chains and dimers.
The four NN interactions form distorted tetrahedral spin chains. 
Magnetic frustration exists in each distorted tetrahedral spin chain. 
We also consider interchain interactions ($J_{ac}$ and $J_a$ interactions) 
shown in Fig. 1(b). 
As will be described later, 
we infer that the space group above 
the transition temperature $T_{\rm N} = 7.9$ K 
is monoclinic $P2_1/m$ (No. 11), although 
deviations from the average symmetry ($Pnma$) are very small.    
In the crystal structure with $P2_1/m$, 
corner ($\alpha$) and center ($\beta$) distorted tetrahedral spin chains 
are inequivalent. 

\begin{figure}
\begin{center}
\includegraphics[width=8cm]{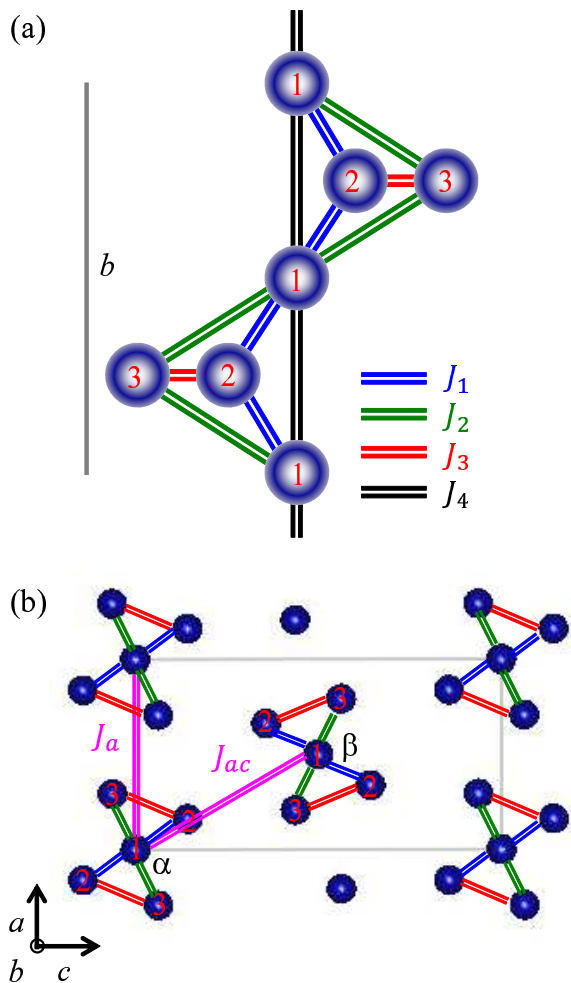}
\caption{
(Color online)
The spin system in Cu$_3$Mo$_2$O$_9$. 
Here, we assume that 
the space group is orthorhombic $Pnma$ (No. 62) 
\cite{Steiner97,Reichelt05}.
There are three crystallographic Cu$^{2+}$ sites having spin-1/2. 
Gray lines correspond to lattice constants. 
(a)
A distorted tetrahedral spin chain parallel to the $b$ axis formed by 
the four types of nearest-neighbor exchange interactions. 
The $J_4$ interaction forms AF chains of Cu1 spins. 
The $J_3$ interaction forms AF dimers of Cu2 and Cu3 spins. 
The $J_1$ and $J_2$ interactions connect the chains and dimers.
(b)
Corner ($\alpha$) and center ($\beta$) distorted tetrahedral spin chains 
projected on the $ac$ plane. 
To explain dispersion relations of magnetic excitations, 
we consider $J_{ac}$ and $J_a$ interchain interactions 
in addition to $J_1 \sim J_4$ interactions.  
We infer that the space group above $T_{\rm N} = 7.9$ K 
is monoclinic $P2_1/m$ (No. 11), although 
deviations from the average symmetry ($Pnma$) are very small. 
The $a$, $b$, and $c$ axes in $Pnma$ correspond to 
the $a$, $b$, and $c$ axes in $P2_1/m$, respectively.  
In the crystal structure with $P2_1/m$, 
$\alpha$ and $\beta$ distorted tetrahedral spin chains 
are inequivalent. 
}
\end{center}
\end{figure}

We determined dispersion relations of magnetic excitations 
from inelastic neutron scattering results \cite{Kuroe11a} of a single crystal 
synthesized using the continuous solid-state 
crystallization method \cite{Oka11}. 
We observed hybridization of magnetic excitations of the chains and dimers.  
The dispersion relations can be explained using the above spin model 
with $J_4 = 4.0$ meV, $J_3 =5 .8$ meV, $J_1 \sim 1$ meV, $J_2 \sim 1$ meV, 
$J_{ac}=0.19$ meV (AF), and $J_{a}=-0.19$ meV (ferromagnetic) 
based on a chain mean-field theory with random phase approximation 
\cite{Kuroe11a,Essler97,Kenzelmann01,Zheludev03}
or 
with $J_4 = 6.50$ meV, $J_3 =5 .70$ meV, $|J_1-J_2| = 3.06$ meV, 
$J_{ac}=0.04$ meV, and $J_{a}=-0.04$ meV 
based on a spin-wave theory \cite{Matsumoto12}.

Cu$_3$Mo$_2$O$_9$ exhibits 
both AF long-range order and electric polarization 
below $T_{\rm N} = 7.9$ K, 
meaning that 
Cu$_3$Mo$_2$O$_9$ is a multiferroic substance \cite{Kuroe11b}. 
Weak ferromagnetism appears 
only in finite magnetic fields parallel to the $a$ or $c$ axis \cite{Hamasaki08}.
The weak ferromagnetism disappears 
by slight substitution of Zn (0.5 \%) for Cu sites \cite{Hase08}.
The electric polarization appears 
parallel to the $c$ axis at zero magnetic field and 
becomes parallel to the $a$ axis 
in the magnetic field parallel to the $c$ axis 
with $H > 8$ T. 
Several magnetic-field induced phase transitions and 
2/3 magnetization plateau appear 
in the magnetization curves up to 74 T 
\cite{Hamasaki09,Kuroe14b}. 
The magnetic and dielectric properties change simultaneously 
at the transition fields. 
Temperature - magnetic field phase diagrams 
have been obtained \cite{Kuroe14b}. 
Raman scattering experiments were performed \cite{Sato14}. 
The Raman scattering spectra did not change at $T_{\rm N}$.  
Therefore, change of the crystal structure caused by the multiferroic transition
is negligibly small. 

It is important to determine the magnetic structure of 
Cu$_3$Mo$_2$O$_9$ to understand the above-mentioned results.
Another group performed neutron powder diffraction experiments and 
reported the magnetic structure of Cu$_3$Mo$_2$O$_9$ \cite{Vilminot09}.
Ferromagnetic behavior parallel to the $b$ axis is expected in the result, 
whereas the ferromagnetic behavior parallel to the $a$ or $c$ axis
was observed in our magnetization results of single crystals \cite{Hamasaki08}. 
We considered that 
reinvestigation of the magnetic structure is necessary. 
Accordingly, we performed neutron powder diffraction experiments. 

\section{Experimental Methods}

We synthesized Cu$_3$Mo$_2$O$_9$ powder 
using a solid-state-reaction method 
at 1023 K in air for 150 h with intermediate grindings. 
We performed X-ray powder diffraction experiments at room temperature 
using a RIGAKU RINT-TTR III diffractometer.
The wavelength $\lambda$ is 1.542~\AA \ (Cu $K_{\alpha}$).
We carried out neutron powder diffraction experiments 
at the Swiss spallation neutron source (SINQ) 
at Paul Scherrer Institute (PSI).  
We used the high-resolution powder diffractometer for thermal neutrons (HRPT) 
in the high intensity mode ($\Delta d/d\geq1.8\times 10^{-3}$) \cite{hrpt} 
and 
the high-intensity cold neutron powder diffractometer (DMC). 
We used neutrons with $\lambda = 1.886$ and 4.507~\AA 
\ at HRPT and DMC diffractometers, respectively.  
Powder was filled into a vanadium container 
with 8 mm diameter and 55 mm height for the HRPT experiments 
and that 
with 10 mm diameter and 55 mm height for the DMC experiments. 

We performed Rietveld refinements of the crystal and magnetic structures
using the {\tt FULLPROF Suite}  program package~\cite{Rodriguez93}  
with its internal tables for scattering lengths and magnetic form factors. 
Symmetry analysis was done  by using {\tt ISODISTORT} tool based on
{\tt ISOTROPY} software \cite{isot,isod} and {\tt BASIREP} program of  {\tt FULLPROF Suite}.

\section{Results and Analyses}

\subsection{Reinvestigation into the crystal structure above $T_{\rm N}$}

The space group of Cu$_3$Mo$_2$O$_9$ at room temperature has been
reported to be orthorhombic $Pnma$ (No. 62)
\cite{Steiner97,Reichelt05}. 
As will shown later, 
the symmetry of the crystal structure is important to 
consider weak ferromagnetism and electric polarization. 
Therefore, we reinvestigated the crystal structure above $T_{\rm N}$. 
We measured high statistics x-ray powder diffraction patterns 
at 290 K to investigate whether 
reflections forbidden in $Pnma$ were present or not.  
As shown in Fig. 2, we found five weak reflections 
that were forbidden in $Pnma$. 
We estimated the $d$ values of the five reflections as 
14.8(2), 7.59(2), 5.14(3), 5.02(2), and 2.936(3) \AA \ 
from lower to higher $2 \theta$.  
Assuming an orthorhombic crystal structure, 
we calculated the lattice constants as  
$a = 7.685(2), b = 6.878(1)$, and $c = 14.650(3)$ \AA \ from 
an x-ray diffraction pattern at 290 K between $2 \theta = 5$ and $80^{\circ}$. 
We obtained the $d$ values as 
14.650(3), 7.685(2), 5.125(2), 5.014(2), and 2.930(1) \AA \ 
for forbidden reflections at 
$(0 0 1)$, $(1 0 0)$, $(1 1 0)$, $(0 1 2)$, and $(0 0 5)$. 
We consider that 
we can assign the indices to the five reflections.  
The diffraction results suggest that 
the space group at 290 K could have a lower symmetry than $Pnma$.
In maximal subgroups of $Pnma$, 
only monoclinic $P2_1/m$ (No. 11) can have all the forbidden reflections. 
The $a$, $b$, and $c$ axes in $Pnma$ correspond to 
the $a$, $b$, and $c$ axes in $P2_1/m$, respectively.  
We observed no phase transition above $T_{\rm N}$ 
in the specific heat of Cu$_3$Mo$_2$O$_9$ \cite{Hamasaki08}.  
Consequently, we tentatively suggest that 
the space group above $T_{\rm N}$ is $P2_1/m$.  
The five reflections forbidden in $Pnma$ are very weak. 
For example, the forbidden $(0 0 5)$ reflection is
weaker than the weak allowed $(1 2 2)$ reflection shown in Fig. 2(d),
which is only 0.72~\% of the largest Bragg reflection at $(0 2 0)$.
Accordingly, 
deviations from the average symmetry are very small and  
we cannot rigorously prove that 
the space group at 290 K is $P2_1/m$.

\begin{figure}
\begin{center}
\includegraphics[width=8cm]{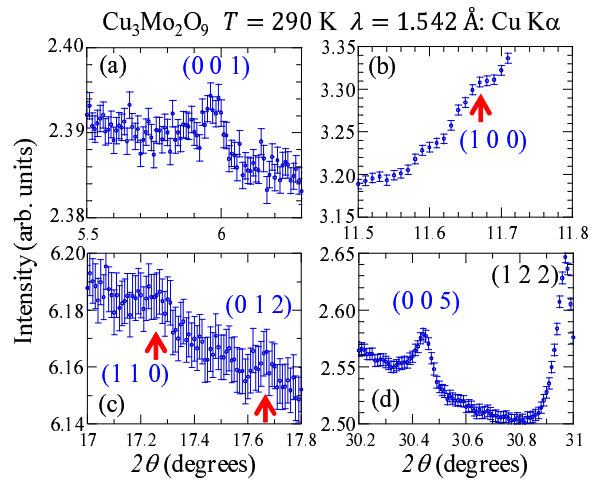}
\caption{ 
(Color online) 
X-ray powder diffraction patterns of
Cu$_3$Mo$_2$O$_9$ at 290 K measured using 
a RIGAKU RINT-TTR III diffractometer 
($\lambda = 1.542$ \AA, Cu $K_{\alpha}$).  
Five reflections forbidden in $Pnma$ were observed at 
$(0 0 1)$, $(1 0 0)$, $(1 1 0)$, $(0 1 2)$, and $(0 0 5)$.  
}
\end{center}
\end{figure}

The circles in Fig. 3 show a neutron powder diffraction pattern 
of Cu$_3$Mo$_2$O$_9$ at 12 K just above $T_{\rm N} = 7.9$~K
recorded using the HRPT diffractometer with $\lambda = 1.886$~\AA. 
We performed Rietveld refinements using $Pnma$ 
to evaluate crystal structure parameters at 12 K. 
The line on the experimental pattern 
indicates the result of Rietveld refinements.  
The line agrees well with the experimental pattern. 
The refined crystal structure parameters are presented in Table I. 
We performed symmetry analysis and Rietveld refinements 
based on both $P2_1/m$ and $Pnma$ to
determine the magnetic structure.  
We expected that atomic positions determined using $P2_1/m$ 
would be almost the same as those determined using $Pnma$
because the forbidden reflections are very weak.  
Therefore, we used
the values in Table I for the following Rietveld refinements to
determine the magnetic structure.  
We describe below 
the relation between the positions in $Pnma$ and $P2_1/m$.  
The basis transformation is given by identity matrix without origin shift.  
The Cu1 ($4a$) position in $Pnma$ is split into 
Cu11 ($2a$) and Cu12 ($2d$) positions in $P2_1/m$. 
The Cu2 and Cu3 ($4c$) positions are split into 
Cu21 and Cu22 ($2e$)-, and Cu31 and Cu32 ($2e$)-positions, respectively.  
In $P2_1/m$, the corner ($\alpha$) and center ($\beta$)
distorted tetrahedral spin chains are inequivalent.

\begin{figure}
\begin{center}
\includegraphics[width=8cm]{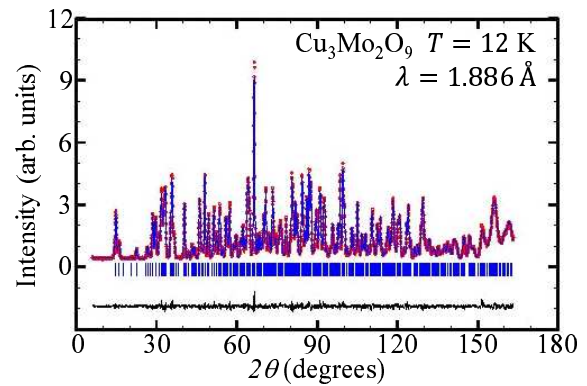}
\caption{
(Color online)
A neutron powder diffraction pattern of Cu$_3$Mo$_2$O$_9$ 
at 12 K (above the transition temperature $T_{\rm N} = 7.9$ K)
measured using the HRPT diffractometer ($\lambda = 1.886$ \AA). 
Lines on the observed pattern and at the bottom 
show a Rietveld refined pattern obtained using 
the crystal structure with $Pnma$ (No. 62)
and the difference between the observed and 
the Rietveld refined patterns, respectively. 
Hash marks represent positions of nuclear reflections.
}
\end{center}
\end{figure}

\begin{table*}
\caption{\label{table1}
Structural parameters of Cu$_3$Mo$_2$O$_9$ derived from 
Rietveld refinements of
the HRPT neutron powder diffraction pattern at 12 K. 
We used orthorhombic $Pnma$ (No. 62).  
The lattice constants at 12 K are
$a=7.629(1)$ \AA, $b=6.876(1)$ \AA, and $c=14.573(1)$ \AA. 
Estimated standard deviations are shown in parentheses. 
The reliability factors of the refinement are 
$R_{\rm p}=3.26$~\%, $R_{\rm wp}=4.16$~\%, $R_{\rm exp}=2.75$~\%, 
and $\chi^2=2.29$. 
}
\begin{ruledtabular}
\begin{tabular}{cccccc}
Atom & Site & $x$ & $y$ & $z$ & $B_{\rm iso}$ \AA$^2$  \\
\hline
Cu1 & 4{\it a} & 0 & 0 & 0 & 0.23(3)\\
Cu2 & 4{\it c} & 0.1620(2) & 0.75 & 0.1430(1) & 0.09(3)\\
Cu3 & 4{\it c} & 0.2064(2) & 0.25 & 0.4373(1) & 0.14(3)\\ 
Mo1 & 4{\it c} & 0.2657(3) & 0.25 & 0.1696(1) & 0.18(3)\\ 
Mo2 & 4{\it c} & 0.1547(3) & 0.75 & 0.3885(1) & 0.19(4)\\
 O1 & 4{\it c} & 0.0875(3) & 0.75 & 0.0159(1) & 0.22(5)\\       
 O2 & 4{\it c} & 0.2075(3) & 0.75 & 0.2715(2) & 0.32(4)\\       
 O3 & 4{\it c} & 0.4310(3) & 0.75 & 0.0969(2) & 0.27(5)\\       
 O4 & 8{\it d} & 0.2481(2) & 0.9636(2) & 0.4397(1) & 0.31(3)\\       
 O5 & 8{\it d} & 0.1410(2) & 0.0369(2) & 0.1373(1) & 0.42(3)\\       
 O6 & 4{\it c} & 0.3029(3) & 0.25 & 0.2857(2) & 0.44(4)\\       
 O7 & 4{\it c} & 0.4692(3) & 0.25 & 0.1103(2) & 0.33(5)\\  
\end{tabular}
\end{ruledtabular}
\end{table*}

\subsection{Magnetic structure}

We measured neutron powder diffraction patterns at 1.6 and 12 K 
(below and above $T_{\rm N} = 7.9$ K, respectively) 
using the DMC diffractometer with $\lambda = 4.507$~\AA.  
Figure 4 shows the difference pattern 
made by subtracting the 12 K diffraction pattern from the 1.6 K one.  
Several magnetic reflections are apparent at 1.6 K.  
All the reflections can be indexed in the chemical cell 
with the propagation vector {\bf k} = {\bf 0}.

\begin{figure}
\begin{center}
\includegraphics[width=8cm]{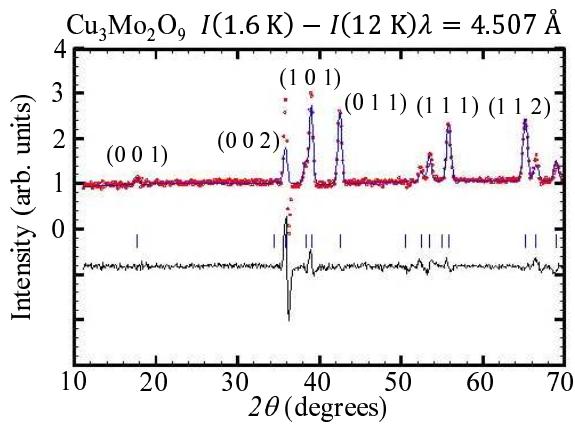}
\caption{
(Color online)
A difference pattern 
made by subtracting a neutron powder diffraction pattern 
of Cu$_3$Mo$_2$O$_9$ at 12 K 
from that at 1.6 K ($T_{\rm N} = 7.9$ K). 
The diffraction patterns were measured 
using the DMC diffractometer ($\lambda = 4.507$ \AA). 
Lines on the observed pattern  
show a Rietveld refined pattern obtained using 
$\Gamma_1^{-}$ ($\tau 2$) in $P2_1/m$. 
Lines at the bottom 
show the difference between the observed and 
the Rietveld refined patterns. 
Hash marks represent positions of magnetic reflections.
We described
indices of several magnetic reflections. 
}
\end{center}
\end{figure}

The magnetic reflections are much weaker than major nuclear reflections.  
Therefore, we have performed the analysis for
the difference pattern shown in Fig. 4 
to determine the magnetic structure of Cu$_3$Mo$_2$O$_9$.  
For {\bf k} = {\bf 0}, 
there are four one dimensional (1D) 
real irreducible representations (IRs) 
that enter the magnetic representation decomposition, 
resulting in four possible Shubnikov groups based on $P2_1/m$.  
The best-fit candidate is $P2_1/m'$, 
which corresponds to $\Gamma_1^{-}$ ($\tau 2$).  
The nomenclature for the IR is given according to Ref.~\cite{isod} with
Kovalev's notation in the parenthesis.  
The $a$ and $c$ components of magnetic moments 
are allowed on the dimer sites ($2e$), 
whereas the magnetic moments on the chain sites ($2a$ and $2d$) 
are zero by symmetry of $\Gamma_1^{-}$.  
There are two IRs even by inversion operator 
$\Gamma_2^{+}$ ($\tau 3$) and $\Gamma_1^{+}$ ($\tau 1$), 
that allow non-zero chain moments.  
We performed Rietveld refinements using
$\Gamma_1^{-}$ for dimer moments and 
$\Gamma_2^{+}$ or $\Gamma_1^{+}$ for chain moments.  
In both cases, 
the reliability factors are smallest for 
the chain moments close to zero.

As shown in Fig. 4, 
the experimental pattern can be explained well 
by the line calculated using only $P2_1/m'$. 
The reliability factors are 
$R_{\rm wp}=4.90$~\%, $R_{\rm exp}=2.85$~\%, and $\chi^2=2.95$. 
The magnetic structure is shown in Fig. 5. 
Magnetic moments on dimer sites are
${\bf M}_{21} = (0.57(2), 0, 0.04(3)) \mu_{\rm B}$ on Cu21 sites, 
${\bf M}_{22} = (0.46(3), 0, -0.50(3)) \mu_{\rm B}$ on Cu22 sites, 
${\bf M}_{31} = (0.27(4), 0, 0.69(3)) \mu_{\rm B}$ on Cu31 sites, and 
${\bf M}_{32} = (0.04(4), 0, -0.50(3)) \mu_{\rm B}$ on Cu32 sites.  
The magnitude of the moments is
$M_{21} = 0.57(2) \mu_{\rm B}$, $M_{22} = 0.69(3) \mu_{\rm B}$,
$M_{31} = 0.74(4) \mu_{\rm B}$, and $M_{32} = 0.50(3) \mu_{\rm B}$.
The angle between two moments 
in Cu21-Cu32 and Cu22-Cu31 dimers are
$90(8)^{\circ}$ and $116(6)^{\circ}$, respectively.  
Two neighboring dimer moments in each distorted tetrahedral spin chain 
are antiparallel to each other as indicated by the blue or green line.  
The chain moments may exist 
but the magnitudes are very small,   
meaning that spins in chains are nearly disordered.
The magnetic structure indicates that 
a partial disordered (PD) state is realized.  
In the previous papers \cite{Hamasaki08,Hase08}, 
we inferred that 
the main component of ordered moments in chains was the $b$ component 
and that
spins in dimers were nearly spin singlet.  
The previous inferences are incorrect.

\begin{figure}
\begin{center}
\includegraphics[width=8cm]{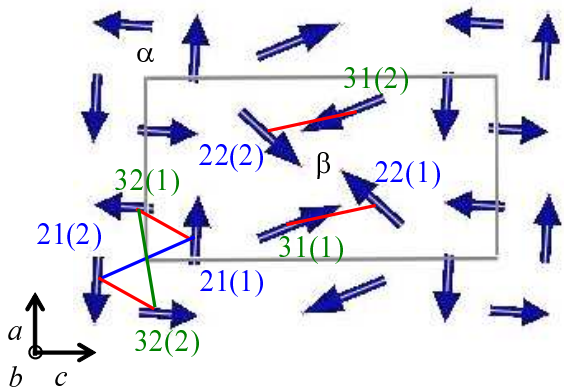}
\caption{
(Color online)
The magnetic structure of Cu$_3$Mo$_2$O$_9$ obtained using 
$\Gamma_1^{-}$ ($\tau 2$) in $P2_1/m$. 
Magnetic moment vectors in dimers are 
${\bf M}_{21} = (0.57(2), 0, 0.04(3)) \mu_{\rm B}$ on Cu21(1) sites,  
${\bf M}_{22} = (0.46(3), 0, -0.50(3)) \mu_{\rm B}$ on Cu22(1) sites, 
${\bf M}_{31} = (0.27(4), 0, 0.69(3)) \mu_{\rm B}$ on Cu31(1) sites,  and 
${\bf M}_{32} = (0.04(4), 0, -0.50(3)) \mu_{\rm B}$ on Cu32(1) sites. 
Symmetry operators of moments on $2e$ sites (dimer moments)
for $\Gamma_1^{-}$ ($\tau 2$) are 
(1) $[u, 0, w]$ and  
(2) $[\bar u, 0, \bar w]$. 
The chain moments are very small if they exist. 
}
\end{center}
\end{figure}


We investigated the magnetic structure assuming $Pnma$  
that has been reported as the space group at room temperature. 
As described below, 
the obtained magnetic structures are similar to that in Fig. 5. 
In the case of ${\bf k} = {\bf 0}$, 
there are eight 1D real IRs, 
resulting in eight Shubnikov groups based on $Pnma$.  
The best-fit candidate is $Pnm'a$, 
which corresponds to $\Gamma_4^{-}$ ($\tau 4$).  
The $a$ and $c$ components of ordered magnetic moments
can be finite on dimer sites ($4c$), 
whereas no ordered magnetic moment on chain sites ($4a$) 
is allowed by symmetry of $\Gamma_4^{-}$.
In addition, 
the $(0 0 1)$ magnetic reflection observed at $2 \theta = 17.7^{\circ}$ 
is forbidden in $\Gamma_4^{-}$.  
There are four IRs even by inversion operator 
that enter the magnetic representation decomposition for the chain moments.
Among the four, the $(0 0 1)$ magnetic reflection is allowed in 
$\Gamma_2^{+}$ ($\tau 7$) and $\Gamma_1^{+}$ ($\tau 1$). 
In refinements using 
$\Gamma_4^{-}$ and $\Gamma_2^{+}$ for 
dimer and chain moments, respectively, 
we obtained magnetic moments as 
${\bf M}_1 = (0.085(9), 0, 0) \mu_{\rm B}$ on Cu1 sites,
${\bf M}_2 = (0.555(8), 0, 0.283(6)) \mu_{\rm B}$ on Cu2 sites, and
${\bf M}_3 = (0.132(7), 0, 0.604(6)) \mu_{\rm B}$ on Cu3 sites.  
The reliability factors are
$R_{\rm wp}=5.02$~\%, $R_{\rm exp}=2.86$~\%, and $\chi^2=3.08$.  
The magnitude of the Cu2 and Cu3 moments is $0.62(1) \mu_{\rm B}$.  
The angle between two moments in each dimer is $105(2)^{\circ}$.  
In refinements using 
$\Gamma_4^{-}$ and $\Gamma_1^{+}$ for 
dimer and chain moments, respectively, 
we obtained magnetic moments as 
${\bf M}_1 = (0, 0.085(9), 0.13(2)) \mu_{\rm B}$ on Cu1 sites 
${\bf M}_2 = (0.543(9), 0, 0.289(7)) \mu_{\rm B}$ on Cu2 sites, and 
${\bf M}_3 = (0.141(7), 0, 0.593(8)) \mu_{\rm B}$ on Cu3 sites.  
The reliability factors are 
$R_{\rm wp}=4.97$~\%, $R_{\rm exp}=2.86$~\%, and $\chi^2=3.02$. 
The magnitude of the Cu1, Cu2,  and Cu3 moments is 
$0.16(2)$, $0.62(1)$, and $0.61(1) \mu_{\rm B}$, respectively.  
The angle between two moments in each dimer is $105(2)^{\circ}$.  

\section{Discussion}

\subsection{the origin of the magnetic structure}

We consider the origin of the magnetic structure in Cu$_3$Mo$_2$O$_9$. 
To avoid complicated explanations,  
we use Cu1, Cu2, and Cu3 to indicate Cu sites. 
As described, 
the four types of NN interactions ($J_1 \sim J_4$ interactions) shown in Fig. 1 
generate magnetic frustration. 
Six NN interactions influence each Cu1 spin, whereas 
only three NN interactions influence each Cu2 or Cu3 spin.
Probably, the frustration is more effective for Cu1 spins 
than Cu2 and Cu3 spins.
Therefore, Cu1 spins are nearly disordered, whereas 
Cu2 and Cu3 spins can be ordered. 
Two neighboring Cu2 (Cu3) moments in each distorted tetrahedral spin chain 
are antiparallel to each other 
as indicated by the blue (green) line in Fig. 5. 
The internal magnetic fields generated by Cu2 or Cu3 ordered moments 
are canceled out on Cu1 sites in the magnetic structure. 
Therefore, classical magnetic energy is independent of 
the direction of Cu1 moments in the absence of magnetic anisotropy. 
The magnetic structure of the Cu2 and Cu3 moments 
allow the nearly disordered state of Cu1 spins. 
 
A similar PD state is seen in 
the frustrated three-leg-ladder Heisenberg antiferromagnets 
Cu$_3$(OH)$_4${\it A}O$_4$ ({\it A} = S or Se) \cite{Vilminot03,Vilminot07}. 
Cu moments within the inner leg remain random, whereas 
Cu moments within the two outer legs are ordered. 
Six NN interactions influence each inner Cu spin, whereas 
four NN interactions influence each outer Cu spin. 
The frustration may be more effective for inner Cu spins 
than outer Cu spins as expected in Cu$_3$Mo$_2$O$_9$. 
In addition to Cu$_3$(OH)$_4${\it A}O$_4$ ({\it A} = S or Se), 
a part of spins are perfectly or nearly disordered in the ordered state 
in several frustrated Heisenberg antiferromagnets such as  
the spinel antiferromagnet 
GeNi$_2$O$_4$ \cite{Matsuda08}, 
the pyrochlore antiferromagnet Gd$_2$Ti$_2$O$_7$, \cite{Stewart04} and
the triangular lattice antiferromagnets 
CuFeO$_2$ \cite{Mekata93} and Ag$_2$CrO$_2$ \cite{Matsuda12}.  
Exchange interactions in long $M - M$ bonds, 
where $M$ represents a magnetic ion, 
play an important role in the occurrence of the PD states. 

Table II shows possible Shubnikov magnetic groups 
appearing as a result of mixing of two IRs.
Among them, 
bond alternation of the $J_4$ interaction is possible in 
$Pm$ (No. 6) and $Pmc2_1$  (No. 26).
The bond alternation of the $J_4$ interaction may be 
the origin of the very small Cu1 moment. 
Bond alternation of NN interactions 
generates a gapped spin-singlet ground state in an AF spin chain 
as in the spin-Peierls system.
The values of $S^{\rm T}$ and $S^{\rm T}_z$ 
are 0 in a spin-singlet ground state. 
Here, $S^{\rm T}$ and $S^{\rm T}_z$ represent  
a value and a $z$ value, respectively, 
of the sum of spin operators. 
Other $S^{\rm T}_z  = 0$ states can be hybridized with 
the spin-singlet ground state 
of an AF alternating spin chain 
by other interactions \cite{Masuda09}.  
States with $S^{\rm T} > 0$ and $S^{\rm T}_z = 0$ are magnetic. 
For example, $S^{\rm T}_z  = 0$ is zero 
in a collinear two-sublattice AF ordered state, 
although the state is not an eigenstate of the AF Heisenberg models. 
As a result, the ground state of the AF alternating spin chain 
can be magnetic by the hybridization of plural $S^{\rm T}_z  = 0$ states. 
When AF long-range order occurs, 
the magnitude of the ordered moments is expected to be larger 
for a smaller spin gap. 
The alternation ratio is probably very close to 1 and 
a spin gap is very small. 
The spin gap of Cu1 spins is less than 0.2 meV 
in inelastic neutron scattering results of Cu$_3$Mo$_2$O$_9$.  
Accordingly, the bond alternation of the $J_4$ interaction is not 
the origin of the very small Cu1 moment.   

\begin{table*}
\caption{\label{table2}
Possible ferromagnetic moment (FM) and electric polarization (EP) 
in the Shubnikov groups obtained by mixing of two IRs of 
the parent space groups. 
The columns indicate 
A: the parent space group to be considered initially, 
B: the two mixed IRs, 
C: the Shubnikov group obtained by the mixing 
  with the number in BNS settings \cite{isod}, and 
D and E: the direction of the possible FM and EP, respectively,
  in the basis of the parent space group of the column A  
  (the Shubnikov group of the column C). 
The symbol "-" means 
no FM or EP.
The third and sixth lines are added to indicate that 
FM and EP
are impossible in Shubnikov group based on single IR.  
}
\begin{ruledtabular}
\begin{tabular}{ccccc}
A & B & C & D: FM & E: EP\\
\hline
$P2_1/m$ & $\Gamma_1^{-}$, $\Gamma_2^{+}$ & $Pm'$ (6.20) & 
$\perp b$ ($\perp b$) & $\perp b$ ($\perp b$)\\
& $\Gamma_1^{-}$, $\Gamma_1^{+}$ & $P2_1$ (4.7) &
$\parallel b$ ($\parallel b$) & $\parallel b$ ($\parallel b$)\\
& $\Gamma_1^{-}$ & $P2_1/m'$ (11.53) &
- (-) &- (-)\\
$Pnma$ & $\Gamma_4^{-}$, $\Gamma_2^{+}$ & $Pm'c2_1'$ (26.68) &
$\parallel c$ ($\parallel b$) & $\parallel a$ ($\parallel c$)\\
& $\Gamma_4^{-}$, $\Gamma_1^{+}$ & $Pna2_1$ (33.144) &
- (-) & $\parallel b$ ($\parallel c$)\\
& $\Gamma_4^{-}$ & $Pnm'a$ (62.444) &
- (-) &- (-)\\
\end{tabular}
\end{ruledtabular}
\end{table*}

The moments in AF dimers are reduced 
from a classical value ($\sim 1 \mu_{\rm B}$). 
Probably, the magnetic frustration generates 
the reduction of the moments. 
In addition, 
character of a spin-singlet pair may be overlapped and 
may cause the reduction of the moments \cite{Tomiyasu11}. 
We do not understand the reason that 
the angle between two moments in AF dimers 
($90(8)^{\circ}$ or $116(6)^{\circ}$)
is different from $180^{\circ}$ (antiparallel configuration). 
As described, 
long-distance interactions exist in 
several frustrated Heisenberg antiferromagnets showing a PD state.  
Long-distance interactions may exist in Cu$_3$Mo$_2$O$_9$ and 
may affect determination of the angle between two moments in AF dimers. 
The 2/3 magnetization plateau appears in the magnetization curves of 
Cu$_3$Mo$_2$O$_9$ \cite{Kuroe14b}. 
In a Heisenberg model with bilinear and biquadratic terms 
[${\bf S}_i \cdot {\bf S}_j$ and $({\bf S}_i \cdot {\bf S}_j)^2$ terms, respectively], 
magnetization plateau can appear as in CdCr$_2$O$_4$ \cite{Penc04}. 
We are seeking for a model that can explain 
the magnetic excitations, the magnetization curves, and the magnetic structure. 

\subsection{consideration to other experimental results}


We comment on the signs of the exchange interactions 
between spins on chain sites. 
From the dispersion relations of the magnetic excitations, 
the $J_4$ interaction forming Cu1 spin chains parallel to the $b$ axis and 
the interchain $J_{ac}$ interaction are antiferromagnetic, whereas 
the interchain $J_{a}$ interaction is ferromagnetic \cite{Kuroe11a,Matsumoto12}. 
The signs of the interactions agree with 
the signs of the $b$ components  
on $2a$ and $2d$ sites for $\Gamma_2^{+}$ in $P2_1/m$ and 
the signs of the $a$ and $c$ components  
on $2a$ and $2d$ sites for $\Gamma_1^{+}$ in $P2_1/m$, 
although the chain moments are nearly disordered. 
Spin chains can show dispersive magnetic excitations 
even in the absence of magnetic long-range order. 
The disordered component 
generates the magnetic excitations of the chains. 


We consider 
the weak ferromagnetism 
in magnetic fields parallel to the $a$ or $c$ axis 
\cite{Hamasaki08} and 
the electric polarization parallel to the $c$ axis 
at zero magnetic field \cite{Kuroe11b}
below $T_{\rm N} = 7.9$ K in Cu$_3$Mo$_2$O$_9$. 
As described in Table II, 
if the IR for the chain moments is $\Gamma_2^{+}$ in $P2_1/m$, 
the Shubnikov group for the magnetic structure 
is $Pm'$ from the mixing of 
$\Gamma_1^{-}$ and $\Gamma_2^{+}$. 
Dimer sites are $1a$ or $1b$ sites. 
Therefore, all the dimer sites are independent. 
Symmetry operators of chain moments on $2a$ and $2d$ sites are 
(1) $[u, v, w]$ and (2) $[u, \bar v, w]$. 
The weak ferromagnetism perpendicular to the $b$ axis 
is possible in $Pm'$. 
The electric polarization perpendicular to the $b$ axis is possible 
in the crystal structure with the space group $Pm$. 
Considering the basis transformation, 
the $b$ axis in $Pm$ corresponds to the $b$ axis in $P2_1/m$.  
Consequently, 
the experimental results of 
the weak ferromagnetism and electric polarization
may be explained by 
the $Pm'$ model described in the first line in Table II. 
The other models in Table II, on the other hand, 
cannot explain the experimental results. 


We consider that the origin of the weak ferromagnetism
is the dimer moments because of the following reasons. 
The chain moments are very small if they exist. 
It may be doubtful that 
the chain moments can have the observable weak ferromagnetism. 
The weak ferromagnetism disappears 
by substitution of Zn for Cu sites \cite{Hase08}. 
We performed neutron powder diffraction experiments on 
Cu$_{2.88}$Zn$_{0.12}$Mo$_2$O$_9$ and 
Cu$_{2.85}$Zn$_{0.15}$Mo$_2$O$_9$ 
using the HRPT diffractometer.
We observed a few magnetic reflections.   
Probably, the propagation vector is {\bf k} = {\bf 0} 
as in Cu$_3$Mo$_2$O$_9$.
We evaluated crystal structure parameters
above $T_{\rm N}$ assuming $Pnma$.  
Most of Zn ions enter Cu3 sites, 
the remaining Zn ions enter Cu2 sites, and 
almost no Zn ions enter Cu1 sites. 
Influence of Zn substitution is more effective for 
Cu2 and Cu3 sites than for Cu1 sites. 
The results suggest that 
Cu2 and Cu3 moments (dimer moments) 
are responsible for the weak ferromagnetism.  
We will determine the magnetic structure of 
Cu$_{3-x}$Zn$_x$Mo$_2$O$_9$ 
to investigate the origin of the weak ferromagnetism and 
the reason that   
the weak ferromagnetism disappears 
by substitution of Zn for Cu sites.

\section{Conclusion}

We investigated the crystal and magnetic structures of  
the spin-1/2 frustrated antiferromagnet Cu$_3$Mo$_2$O$_9$ 
in which the spin system consists of  
antiferromagnetic chains and dimers.  
The space group at room temperature 
has been reported to be orthorhombic $Pnma$ (No. 62). 
We observed reflections forbidden in $Pnma$ 
in x-ray powder diffraction experiments at room temperature. 
We infer that 
the space group above $T_{\rm N} = 7.9$ K is monoclinic $P2_1/m$  (No. 11)
that is one of maximal subgroups of $Pnma$. 
We determined the magnetic structure of Cu$_3$Mo$_2$O$_9$
in neutron powder diffraction experiments.  
The experimental diffraction pattern 
can be explained well by the line calculated 
using the irreducible representation (IR) $\Gamma_1^{-}$ ($\tau 2$) 
in $P2_1/m$. 
Magnetic moments on dimer sites lie in the $ac$ planes. 
The magnitudes are $0.50 \sim 0.74 \mu_{\rm B}$.
The angle between two moments in dimers 
is $90(8)^{\circ}$ or $116(6)^{\circ}$. 
The moments on chain sites may exist 
but the magnitudes are very small. 
The magnetic structure indicates that 
a partial disordered state is realized.  
Probably, magnetic frustration influences the magnetic structure. 
If the IR for the chain moments is $\Gamma_2^{+}$ ($\tau 3$) in $P2_1/m$, 
the Shubnikov group for the magnetic structure 
is $Pm'$ from the mixing of 
$\Gamma_1^{-}$ and $\Gamma_2^{+}$. 
The weak ferromagnetism perpendicular to the $b$ axis 
is possible in $Pm'$. 
The electric polarization perpendicular to the $b$ axis is possible 
in the crystal structure with the space group $Pm$ (No. 6). 
The $b$ axis in $Pm$ corresponds to the $b$ axis in $P2_1/m$ and $Pnma$. 
Consequently, 
the experimental results of 
the weak ferromagnetism and electric polarization
may be explained by the $Pm'$ model. 

\begin{acknowledgments}

We are grateful to 
H. Eisaki, T. Goto, T. Ito, K. Kaneko, Y. Kawamura, R. Kiyanagi, M. Kohno, 
T. Masuda, M. Matsuda, M. Matsumoto, S. Matsumoto, 
Y. Noda, K. Oka, T. Suzuki, K. Tomiyasu, and O. Zaharko 
for fruitful discussion or experimental supports. 
The neutron powder diffraction experiments were performed 
by using the DMC and HRPT diffractometers 
at SINQ, PSI, Switzerland 
(proposal nos. 20131441 and 20131464, respectively).
The experiments were transferred from 
5G:PONTA at JRR-3 with the approval of Institute for Solid State Physics, 
The University of Tokyo (proposal nos. 14806 and 14807) and  
Japan Atomic Energy Agency, Tokai, Japan.
This work was partially supported by grants from NIMS. 

\end{acknowledgments}

\newpage 

\begin{references}

\bibitem{Anderson73}
P. W. Anderson, 
Mater. Res. Bull. {\bf 8}, 153 (1973).

\bibitem{Miyashita84}
S. Miyashita and H. Shiba, 
J. Phys. Soc. Jpn. {\bf 53}, 1145 (1984).

\bibitem{Onoda07}
S. Onoda and N. Nagaosa, 
Phys. Rev. Lett. {\bf 99}, 027206 (2007).

\bibitem{Tsunetsugu06}
H. Tsunetsugu and M. Arikawa, 
J. Phys. Soc. Jpn. {\bf 75}, 083701 (2006).

\bibitem{Zhitomirsky08}
M. E. Zhitomirsky, 
Phys. Rev. B {\bf 78}, 094423 (2008).

\bibitem{Kawamura10}
H. Kawamura, A. Yamamoto, and T. Okubo, 
J. Phys. Soc. Jpn. {\bf 79}, 023701 (2010).

\bibitem{Haldane82a}
F. D. M. Haldane, 
Phys. Rev. B {\bf 25}, 4925 (1982).

\bibitem{Haldane82b}
F. D. M. Haldane, 
Phys. Rev. B {\bf 26}, 5257 (1982).

\bibitem{Okamoto92}
K. Okamoto and K. Nomura, Phys. Lett. A {\bf 169}, 433 (1992).

\bibitem{Hase93a}
M. Hase, I. Terasaki, and K. Uchinokura, 
Phys. Rev. Lett. {\bf 70}, 3651 (1993).

\bibitem{Hase93b}
M. Hase, I. Terasaki, Y. Sasago, K. Uchinokura, and H. Obara, 
Phys. Rev. Lett. {\bf 71}, 4059 (1993).

\bibitem{Hase93c}
M. Hase, I. Terasaki, K. Uchinokura, M. Tokunaga, 
N. Miura, and H. Obara, 
Phys. Rev. B {\bf 48}, 9616 (1993).

\bibitem{Lorenzo94}
J. E. Lorenzo, K. Hirota, G. Shirane, J. M. Tranquada, M. Hase, K. Uchinokura, H. Kojima, I. Tanaka, and Y. Shibuya, Phys. Rev. B {\bf 50}, 1278 (1994).

\bibitem{Castilla95}
G. Castilla, S. Chakravarty, and V. J. Emery, 
Phys. Rev. Lett. {\bf 75}, 1823 (1995).

\bibitem{Riera95}
J. Riera and A. Dobry, Phys. Rev. B {\bf 51}, 16098 (1995).

\bibitem{Tanokura03}
Y. Tanokura, Y. Oono, S. Ikeda, H. Kuroe, T. Sekine, T. Masuda and K. Uchinokura, 
Phys. Rev. B {\bf 68}, 054412 (2003).

\bibitem{Chubukov91}
A. V. Chubukov, 
Phys. Rev. B {\bf 44}, 4693 (1991).

\bibitem{Kolezhuk05}
A. Kolezhuk and T. Vekua, 
Phys. Rev. B {\bf 72}, 094424 (2005).

\bibitem{Heidrich06}
F. Heidrich-Meisner, A. Honecker, and T. Vekua, 
Phys. Rev. B {\bf 74}, 020403(R) (2006).

\bibitem{Vekua07}
T. Vekua, A. Honecker, H.-J. Mikeska, and F. Heidrich-Meisner, 
Phys. Rev. B {\bf 76}, 174420 (2007).

\bibitem{Kecke07}
L. Kecke, T. Momoi, and A. Furusaki, 
Phys. Rev. B {\bf 76}, 060407(R) (2007).

\bibitem{Hikihara08}
T. Hikihara, L. Kecke, T. Momoi, and A. Furusaki, 
Phys. Rev. B {\bf 78}, 144404 (2008).

\bibitem{Sudan09}
J. Sudan, A. Luscher, and A. M. L\"auchli, 
Phys. Rev. {\bf B} 80, 140402(R) (2009).

\bibitem{Sato09}
M. Sato, T. Momoi, and A. Furusaki, 
Phys. Rev. B {\bf 79}, 060406(R) (2009).

\bibitem{Heidrich09}
F. Heidrich-Meisner, I. P. McCulloch, and A. K. Kolezhuk, 
Phys. Rev. B {\bf 80}, 144417 (2009).

\bibitem{Sato11}
M. Sato, T. Hikihara, and T. Momoi, 
Phys. Rev. B {\bf 83}, 064405 (2011).

\bibitem{Hase04}
M. Hase, H. Kuroe, K. Ozawa, O. Suzuki, H. Kitazawa, G. Kido, and T. Sekine, 
Phys. Rev. B {\bf 70}, 104426 (2004).

\bibitem{Drechsler07} 
S.-L. Drechsler, O. Volkova, A. N. Vasiliev, N. Tristan, J. Richter, M. Schmitt, H. Rosner, J. M\'alek, R. Klingeler, A. A. Zvyagin, and B. B\"uchner, 
Phys. Rev. Lett. {\bf 98}, 077202 (2007).

\bibitem{Okamoto99}
K. Okamoto, T. Tonegawa, Y. Takahashi and M. Kaburagi, 
J. Phys. Condens. Matter {\bf 15}, 5979 (1999).

\bibitem{Kikuchi05}
H. Kikuchi, Y. Fujii, M. Chiba, S. Mitsudo, T. Idehara, 
T. Tonegawa, K. Okamoto, T. Sakai, T. Kuwai and H. Ohta, 
Phys. Rev. Lett. {\bf 94}, 227201 (2005).

\bibitem{Hamasaki08}
T. Hamasaki, T. Ide, H. Kuroe, T. Sekine, M. Hase, I. Tsukada, and T. Sakakibara, 
Phys. Rev B {\bf 77}, 134419 (2008).

\bibitem{Steiner97}
U. Steiner and W. Reichelt: Acta Cryst. C {\bf 53}, 1371 (1997).

\bibitem{Reichelt05} 
W. Reichelt, U. Steiner, T. S\"ohnel, O. Oeckler, V. Duppel and L. Kienle, 
Z. Anorg. Allg. Chem. {\bf 631}, 596 (2005).

\bibitem{Kuroe11a}
H. Kuroe, T. Hamasaki, T. Sekine, M. Hase, K. Oka, T. Ito, H. Eisaki, 
K. Kaneko, N. Metoki, M. Matsuda, and K. Kakurai, 
Phys. Rev. B {\bf 83}, 184423 (2011).

\bibitem{Oka11}
K. Oka, T. Ito, H. Eisaki, M. Hase, T. Hamasaki, H. Kuroe, and T. Sekine, 
J. Crystal Growth {\bf 334}, 108 (2011).

\bibitem{Essler97}
F. H. L. Essler, A. M. Tsvelik, and G. Delfino, 
Phys. Rev. B {\bf 56}, 11001 (1997).

\bibitem{Kenzelmann01}
M. Kenzelmann, A. Zheludev, S. Raymond, E. Ressouche, T. Masuda, 
P. B\'oni, K. Kakurai, I. Tsukada, K. Uchinokura, and R. Coldea, 
Phys. Rev. B {\bf 64}, 054422 (2001).

\bibitem{Zheludev03}
A. Zheludev, S. Raymond, L.-P. Regnault, F. H. L. Essler, K. Kakurai, 
T. Masuda, and K. Uchinokura, 
Phys. Rev. B {\bf 67}, 134406 (2003).

\bibitem{Matsumoto12}
M. Matsumoto, H. Kuroe, T. Sekine, and M. Hase, 
J. Phys. Soc. Jpn. {\bf 81}, 024711 (2012).

\bibitem{Kuroe11b}
H. Kuroe, R. Kino, T. Hosaka, M. Suzuki, S. Hachiuma, T. Sekine, M. Hase, 
K. Oka, T. Ito, H. Eisaki, M. Fujisawa, S. Okubo, and H. Ohta, 
J. Phys. Soc. Jpn. {\bf 80}, 083705 (2011).

\bibitem{Hase08}
M. Hase, H. Kitazawa K. Ozawa, T. Hamasaki, H. Kuroe, and T. Sekine, 
J. Phys. Soc. Jpn. {\bf 77}, 034706 (2008).

\bibitem{Hamasaki09}
T. Hamasaki, H. Kuroe, T. Sekine, M. Hase, and H. Kitazawa, 
J. Phys.: Conf. Series {\bf 150}, 042047 (2009).

\bibitem{Kuroe14b}
H. Kuroe, K. Aoki, R. Kino, T. Sato, H. Kuwahara, T. Sekine, 
T. Kihara,Y. Kohama, M. Akaki, 
M. Tokunaga, M. Hase, T. Takehana, H. Kitazawa, K. Oka, T. Ito, and H. Eisaki, 
JPS Conf. Proc. {\bf 3}, 014036 (2014). 

\bibitem{Sato14}
T. Sato, K. Aoki, R. Kino, H. Kuroe, T. Sekine, M. Hase, K. Oka, T. Ito, and H. Eisaki, 
JPS Conf. Proc. {\bf 3}, 014035 (2014).

\bibitem{Vilminot09}
S. Vilminot, G. Andr\'e, and M. Kurmoo,
Inorg. Chem. {\bf 48}, 2687 (2009).

\bibitem{hrpt}
P. Fischer, G. Frey, M. Koch, M. Koennecke, V. Pomjakushin, J. Schefer, R. Thut,
N. Schlumpf, R. Buerge, U. Greuter, S. Bondt, and E. Berruyer, 
Physica B, {\bf 276-278}, 146 (2000); [http://sinq.web.psi.ch/hrpt].

\bibitem{Rodriguez93}
J. Rodriguez-Carvajal, Physica B  {\bf 192}, 55 (1993); 
[http://www.ill.eu/sites/fullprof/]. 

\bibitem{isot}
H. T. Stokes and D. M. Hatch, 
{\it Isotropy Subgroups of the 230 Crystallographic Space Groups},
(World Scientific Publishing, Singapore, 1988).  

\bibitem{isod}
B. J. Campbell, H. T. Stokes, D. E. Tanner, and D. M. Hatch, 
J. Appl. Cryst. {\bf 39}, 607 (2006).


\bibitem{Vilminot03}
S. Vilminot, M. Richard-Plouet, G. Andr\'e, D. Swierczynski, M. Guillot, 
F. Bour\'ee-Vigneron, and M. Drillon, 
J. Solid State Chem. {\bf 170}, 255 (2003).

\bibitem{Vilminot07}
S. Vilminot, G. Andr\'e, F. Bour\'ee-Vigneron, M. Richard-Plouet, and M. Kurmoo,
Inorg. Chem. {\bf 46}, 10079 (2007).

\bibitem{Matsuda08}
M. Matsuda, J.-H. Chung, S. Park, T. J. Sato, K. Matsuno, H. Aruga Katori, 
H. Takagi, K. Kakurai, K. Kamazawa, Y. Tsunoda, I. Kagomiya, 
C. L. Henley, and S.-H. Lee, 
EPL {\bf 82}, 37006 (2008).

\bibitem{Stewart04}
J. R. Stewart, G. Ehlers, A. S. Wills, S. T. Bramwell, and J. S. Gardner,
J. Phys.: Condens. Matter {\bf 16}, L321 (2004).

\bibitem{Mekata93}
M. Mekata, N. Yaguchi, T. Takagi, T. Sugino, S. Mitsuda, H. Yoshizawa, 
N. Hosoito, and T. Shinjo, 
J. Phys. Soc. Jpn. {\bf 62}, 4474 (1993).

\bibitem{Matsuda12}
M. Matsuda, C. de la Cruz, H. Yoshida, M. Isobe, R. S. Fishman, 
Phys. Rev B {\bf 85}, 144407 (2012).

\bibitem{Masuda09}
T. Masuda, K. Kakurai, and A. Zheludev,  
Phys. Rev B {\bf 80}, 180412(R) (2009).

\bibitem{Tomiyasu11}
K. Tomiyasu, M. K. Crawford, D. T. Adroja, P. Manuel, A. Tominaga, S. Hara, 
H. Sato, T. Watanabe, S. I. Ikeda, J. W. Lynn, K. Iwasa, and K. Yamada, 
Phys. Rev B {\bf 84}, 054405 (2011).

\bibitem{Penc04}
K. Penc, N. Shannon, and H. Shiba, 
Phys. Rev. Lett. {\bf 93}, 197203 (2004).

\end{references}

\end{document}